\documentclass[review]{elsarticle}
\usepackage{lineno,hyperref}

\usepackage{graphicx}
\usepackage{tabularx}
\usepackage{listings}
\usepackage{color}
\usepackage{hyperref}
\usepackage{booktabs}
\usepackage{xltabular}
\usepackage{amssymb}
\usepackage[titletoc,title]{appendix}
\usepackage{multirow}
\usepackage[misc,geometry]{ifsym}
\usepackage{footnote}

\newcolumntype{L}[1]{>{\raggedright\arraybackslash\hspace{0pt}}m{#1}}
\global
\global

\modulolinenumbers
\journal{Journal of Systems and Software}
\begin{document}
\begin{frontmatter}

\title{Selenium-Jupiter: A JUnit 5 extension for Selenium WebDriver}
\author{Boni Garc\'ia\corref{mycorrespondingauthor}}
\ead{boni.garcia@uc3m.es}

\author{Carlos Delgado Kloos\corref{}}
\ead{cdk@it.uc3m.es}

\author{Carlos Alario-Hoyos\corref{}}
\ead{calario@it.uc3m.es}

\author{Mario Munoz-Organero\corref{}}
\ead{munozm@it.uc3m.es}

\address{Universidad Carlos III de Madrid}
\address{Avenida de la Universidad 30, 28911 Legan\'es, Spain}

\cortext[mycorrespondingauthor]{Corresponding author}


\begin{abstract}

Selenium WebDriver is a library that allows controlling web browsers (e.g., Chrome, Firefox, etc.) programmatically. It provides a cross-browser programming interface in several languages used primarily to implement end-to-end tests for web applications. JUnit is a popular unit testing framework for Java. Its latest version (i.e., JUnit 5) provides a programming and extension model called Jupiter. This paper presents Selenium-Jupiter, an open-source JUnit 5 extension for Selenium WebDriver. Selenium-Jupiter aims to ease the development of Selenium WebDriver tests thanks to an automated driver management process implemented in conjunction with the Jupiter parameter resolution mechanism. Moreover, Selenium-Jupiter provides seamless integration with Docker, allowing the use of different web browsers in Docker containers out of the box. This feature enables cross-browser testing,  load testing, and troubleshooting (e.g., configurable session recordings). This paper presents an example case in which Selenium-Jupiter is used to evaluate the performance of video conferencing systems based on WebRTC. This example case shows that Selenium-Jupiter can build and maintain the required infrastructure for complex tests effortlessly.

\end{abstract}

\begin{keyword}
Browser automation \sep Automated testing tools \sep End-to-end testing \sep Selenium WebDriver \sep JUnit \sep Docker
\end{keyword}

\end{frontmatter}


\section{Introduction}
\label{int}

Selenium WebDriver\footnote{\url{https://www.selenium.dev/}} is a library that allows controlling programmatically web browsers (such as Chrome, Firefox, Edge, or Opera). This library can be used for multiple purposes related to browser automation (e.g., web scraping). Nevertheless, it is primarily employed to test web applications and is considered the de facto standard framework for end-to-end web testing, supporting a solid automated testing industry worldwide \cite{niranjanamurthy2018functional}.

Selenium WebDriver allows controlling web browsers using different language bindings such as Java, JavaScript, Python, C\#, or Ruby. In end-to-end tests, the Selenium WebDriver Application Programming Interface (API) calls are typically embedded in test cases using a unit testing framework. A recent study of the Selenium ecosystem identifies Java as the preferred language binding and JUnit as the most frequently used testing framework for developing end-to-end tests with Selenium WebDriver \cite{garcia2020survey}. JUnit is one of the most popular testing frameworks for the Java Virtual Machine (JVM), and one of the most influential in software engineering \cite{janzen2006influence}. JUnit 5\footnote{\url{https://junit.org/junit5/}} is the next generation of JUnit, first released in September 2017. JUnit 5 provides a brand-new programming and extension model called Jupiter \cite{garcia2017mastering}.

This paper presents Selenium-Jupiter\footnote{\url{https://github.com/bonigarcia/selenium-jupiter}}, an open-source extension of JUnit 5 for Selenium WebDriver. Selenium-Jupiter aims to ease the development of end-to-end tests on top of the Jupiter programming model. Selenium-Jupiter provides automated mechanisms to manage Selenium drivers (e.g., chromedriver, geckodriver) transparently for developers. Besides, it allows using web browsers in Docker containers effortlessly.

The remainder of this paper is structured as follows. Section \ref{back} provides a summary of the technological background supporting this work. Section \ref{mot}  presents the motivation and objectives of Selenium-Jupiter. Then, Sections \ref{des} and \ref{fea} describe its design and relevant features, respectively. To validate the proposed approach, Section \ref{cas} presents the design and results of an example case in which Selenium-Jupiter is used to carry out end-to-end performance testing of videoconferencing web applications (WebRTC). Next, Section \ref{dis} discusses the main outcomes and limitations of this work. Finally, Section \ref{con} summarizes the main conclusions and possible future work of this piece of research.

\section{Background}
\label{back}

\subsection{Selenium WebDriver}
\label{selweb}

\begin{figure}
\centering
\includegraphics[width=0.99\textwidth]{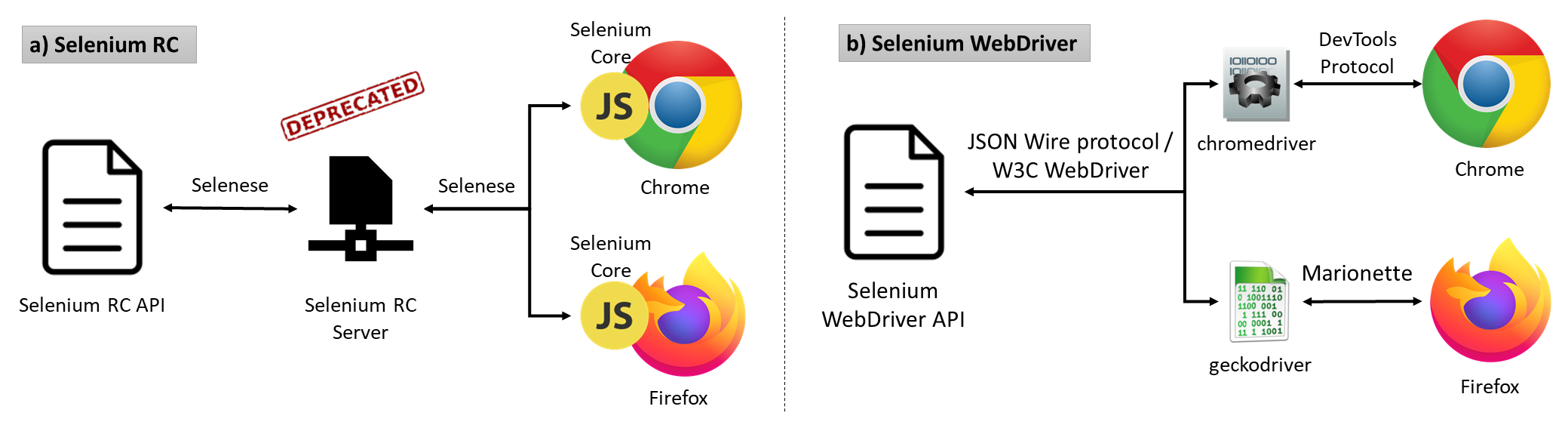}
\caption{Architecture of Selenium RC and Selenium WebDriver}
\label{f1}
\end{figure}

Selenium is an open-source project devoted to providing browser automation. It was first released by Jason Huggins and Paul Hammant in 2004. The first version of Selenium (known nowadays as Selenium Core) was a JavaScript library that interpreted the so-called Selenese commands to impersonate user actions in web applications \cite{bruns2009web}. Huggins and Hammant combined a scripting layer with Selenium Core to create a brand new project called Selenium Remote Control (RC). As shown in Figure \ref{f1}-a, Selenium RC follows a client-server architecture. Clients use a binding language (e.g., Java or JavaScript) to send Selenese commands over HTTP to an intermediate proxy called the Selenium RC server. This proxy injects Selenium Core as a JavaScript library on web browsers launched on-demand, redirecting client requests to Selenium Core \cite{burns2010selenium}. This approach was a pioneer for browser automation at that time. Nevertheless, it suffered from significant limitations. First, and since it is based on JavaScript, several actions cannot be automated (such as mouse movement or headless support, to name a few). Also, Selenium RC introduces a significant overhead that impacts the performance of end-to-end tests.

In parallel, Simon Stewart created a new project called WebDriver in 2007. In the same way that RC, WebDriver allows controlling web browsers using a binding language. Nevertheless, WebDriver is based on the native support of each browser, and therefore, the performance and automation capabilities are superior to those of RC. In 2009, Jason Huggins and Simon Stewart merged Selenium and WebDriver in a new project called Selenium WebDriver (also known as Selenium 2). This way, Selenium RC is nowadays discouraged in favor of Selenium WebDriver \cite{gojare2015analysis}. As shown in Figure \ref{f1}-b, it is necessary to include an intermediate file (the so-called ``driver'') to control a browser with Selenium WebDriver. The driver (e.g., chromedriver for Chrome or geckodriver for Firefox) is a platform-dependent binary file that receives commands from the Selenium WebDriver API and translates them into some browser-specific language. Initially, the communication between the Selenium script and the drivers was done using JSON messages over HTTP in the so-called JSON Wire Protocol \cite{raghavendra2021introduction}. Nowadays, this communication is standardized according to the W3C WebDriver recommendation (also based on JSON messages over HTTP) \cite{stewart2020webdriver}. The JSON Wire Protocol has been discouraged in favor of W3C WebDriver as of Selenium version 4 \cite{garcia2020survey}.

Another component of the Selenium project is Selenium Grid, created by Philippe Hanrigou in 2008. Selenium Grid allows the execution of Selenium WebDriver scripts in remote hosts. To that aim, Selenium Grid comprises a central component called Selenium Hub and a group of nodes. The Hub (also known as Selenium Server) accepts requests from Selenium WebDriver clients and proxies these requests to the nodes. Nodes typically run on multiple operating systems and provide different browsers to be used in WebDriver scripts.

\subsection{JUnit 5}
\label{junit5}

JUnit is a unit testing framework for Java created by Erich Gamma and Kent Beck in 1997 \cite{beck2004junit}. JUnit has been considered as the de facto standard framework for developing unit tests in Java. Thus, the underlying testing model implemented in JUnit has inspired a family of unit testing frameworks in other languages (e.g., C\#, Perl, or Python, to name a few) in the so-called xUnit family \cite{meszaros2007xunit}. As of JUnit 4, the building blocks for developing JUnit test cases are Java annotations. For instance, the annotation \texttt{@Test} is used to identify the methods of a Java class used to exercise and verify a System Under Test (SUT). 

\begin{figure}
\centering
\includegraphics[width=0.9\textwidth]{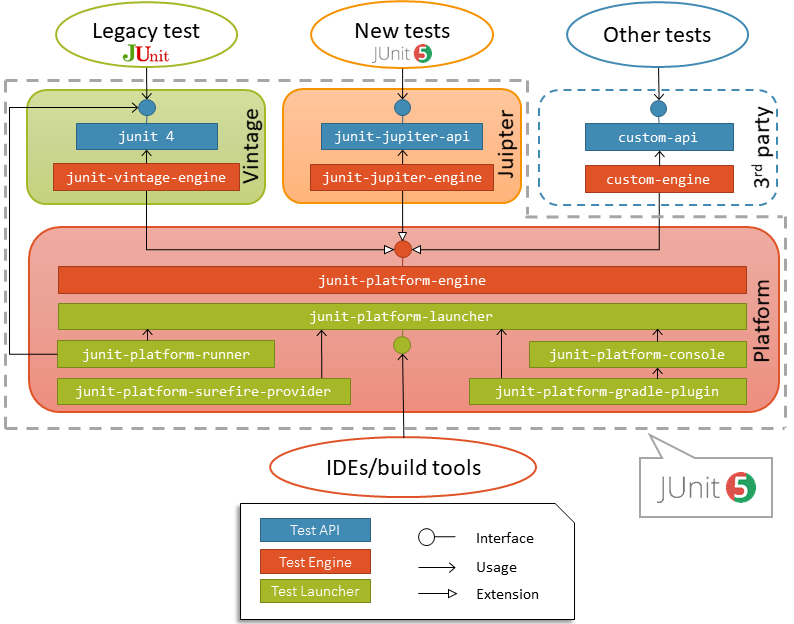}
\caption{Architecture of JUnit 5}
\label{f2}
\end{figure}

Due to some relevant limitations in JUnit 4 (e.g., monolithic architecture or impossibility to compose JUnit runners), a new major version (i.e., JUnit 5) was released in 2017. In this new version of JUnit, Java annotations are again used to declare test cases (i.e., \texttt{@Test} annotation) and also their lifecycle, i.e., the logic executed before (annotations \texttt{@BeforeAll} and \texttt{@BeforeEach}) and after the tests (annotations \texttt{@AfterAll} and \texttt{@AfterEach}). JUnit has been redesigned entirely in version 5, following a modular architecture made up of three components. As shown in Figure \ref{f2}, the first component is called the JUnit Platform, and it is a foundation component to execute tests in the JVM. The second component is called Vintage and provides backward compatibility with legacy JUnit tests (i.e., versions 3 and 4). The third component is called Jupiter and provides a new programming and extension model to execute tests on top of the JUnit Platform \cite{garcia2017mastering}.

The extension model of Jupiter allows adding custom features to the default programming model. To this aim, Jupiter defines an API in which developers can extend different interfaces (called extension points) to provide custom functionality. Table \ref{t1} provides a comprehensive summary of the main features of the Jupiter extension API. Jupiter extensions are registered declaratively using the annotation \texttt{@ExtendWith} or programmatically with  \texttt{@RegisterExtension}.

\begin{table}
\caption{Summary of the Jupiter extension API}
\label{t1}
\begin{tabularx}{\textwidth}{L{1.9cm} L{4.8cm} L{4.8cm}} \\
\hline\noalign{\smallskip}
\textbf{Category} & \textbf{Description} & \textbf{Extension point(s)} \\
\midrule

Test lifecycle callbacks
&
Used to include custom logic in different moments of the test lifecycle
&
\texttt{BeforeAllCallback}, \texttt{BeforeEachCallback}, {\small\texttt{BeforeTestExecutionCallback}}, {\small\texttt{AfterTestExecutionCallback}}, \texttt{AfterEachCallback}, and \texttt{AfterAllCallback}
\\
\midrule

Parameter resolution
&
Used in those extensions that require dependency injection (i.e., parameters injected in test methods or constructors)
&
\texttt{ParameterResolver}
\\
\midrule

Test templates
&
Used to implement \texttt{@TestTemplate} tests (repeated depending on a given context)
&
\texttt{TestTemplateInvocation\-ContextProvider}
\\
\midrule

Conditional test execution
&
Used to enable or disable tests depending on certain conditions
&
\texttt{ExecutionCondition}
\\
\midrule

Exception handling
&
Used to handle exceptions during the test and its lifecycle (i.e., before and after the test)
&
\texttt{TestExecutionException\-Handler}, and \texttt{LifecycleMethodExecution\-ExceptionHandler}
\\
\midrule

Test instance
&
Used to create and process test class instances
&
\texttt{TestInstanceFactory}, \texttt{TestInstancePostProcessor}, and \texttt{TestInstancePre\-DestroyCallback} 
\\
\midrule

Intercepting invocations
&
Used to intercept calls to test code (and decide whether or not these calls proceed)
&
\texttt{InvocationInterceptor}
\\
\midrule

\end{tabularx}
\end{table}

\subsection{Related work}
\label{related}

A recent study about software testing identified Selenium as the most valuable testing framework nowadays, followed by JUnit and Cucumber \cite{cerioli20205}. Thus, Selenium has been used extensively in the literature. For instance, the SmartDriver project \cite{bures2016smartdriver} is a Selenium WebDriver extension based on the separation of the following aspects: 1) Technical aspects associated with the user interface and test logic, and 2) Business concerns related to the SUT. Another approach related to Selenium WebDriver is proposed by Clerissi et al. \cite{clerissi2017towards}, in which Selenium WebDriver tests are automatically generated using textual or UML-based requirements specification. Also, the standard specification to drive browsers automatically by Selenium WebDriver (i.e., the W3C WebDriver recommendation) is used as the foundation to carry out user impersonation as a service, understood as advanced capabilities to build end-to-end tests on the top of cloud infrastructure \cite{garcia2018extending}.

Maintainability and flakiness were identified as the most common problems related to end-to-end testing in a recent survey about the Selenium ecosystem \cite{garcia2020survey}. Software maintainability is the ease with which a software system can be modified \cite{159342}. According to Leotta et al., the adoption of the Page Object Model (POM) might help to make more robust tests while reducing the maintenance efforts significantly \cite{leotta2013improving}. POM is a design pattern in which web pages are modeled using object-oriented classes to minimize code duplication. Another approach for reducing the maintenance costs of Selenium WebDriver tests is related to the proper selection of a web element location strategy (e.g., by identifier or XPath) \cite{leotta2013comparing}. In this arena, the automatic generation of locators based on XPath is proposed to avoid fragility in Selenium \cite{leotta2014reducing}\cite{leotta2015using}. Memon et al. define test flakiness as ``the inability to repeat execution in a reliable manner'' \cite{memon2013automated}. In other words, a test is said to be flaky when its outcomes are non-deterministic (i.e., unreliable under the same conditions) \cite{pinto2020vocabulary}. There are different reasons for flaky tests in Selenium WebDriver reported in the literature, including fragile location strategies \cite{leotta2015using} or incorrect wait strategies (i.e., a configurable timeout for locating web element availability or checking some conditions) \cite{presler2019wait}.

Selenium WebDriver is primarily used to implement functional test cases of web applications. This testing type aims to ensure the correctness of the SUT and detect defects (bugs). Nevertheless, Selenium WebDriver can also be used to verify non-functional requirements, such as performance \cite{garcia2020survey}. In this case, the main objective is to find system bottlenecks or verify that the SUT can operate within the defined boundaries (load testing) \cite{lenka2018performance}. As reported in Section \ref{cas}, this characteristic is used to carry out an example case of end-to-end performance testing of WebRTC applications through Selenium WebDriver.  WebRTC is a set of standard technologies that allow interchanging video and audio in real-time using web browsers \cite{garcia2017webrtc}. The usage of Selenium WebDriver to assess WebRTC applications is a well-known use case in the literature. For instance, it has been used to implement automated tests to evaluate the Quality of Experience (QoE) of WebRTC applications \cite{bertolino2020quality}\cite{garcia2019practical}\cite{garcia2020assessment}. QoE is the degree of satisfaction of the user of an application or service \cite{garcia2019understanding}.

Another relevant technology in this piece of research is Docker. We have witnessed a growing interest in Docker in the last decade since it enables a convenient process for developing and distributing software. Haque et al. present an empirical study identifying practitioners' perspectives on Docker technology by mining Stack Overflow posts \cite{haque2020challenges}. One of the topics under investigation related to application development was web browsers in Docker containers. The findings of this work reveal that developers investigate how to use Selenium WebDriver when using Docker, asking questions like "How do you set up selenium grid using Docker on Windows?" or why WebDriver does not work in a particular platform. Despite its potential utility, there are scarce research efforts to combine Selenium WebDriver and Docker in the literature. One example is ElasTest\footnote{\url{https://elastest.io/}}, an open-source generic and extensible platform supporting end-to-end testing of different application types (including web or mobile, among others) by leveraging the containers technology offered by Docker \cite{bertolino2018testing}. Another example is CAdViSE\footnote{\url{https://github.com/cd-athena/CAdViSE}}, a video streaming framework for load testing of web-based media players using browsers in Docker containers \cite{taraghi2020cadvise}.

\section{Motivation}
\label{mot}

The general objective of this work is to provide a comprehensive programming model for developing end-to-end tests for web applications using Selenium WebDriver on top of JUnit 5. This contribution is entirely missing in the literature to the best of our knowledge and aims to solve several current challenges in end-to-end testing.

As introduced in Section \ref{related}, improving the maintainability and flakiness has been identified as the most relevant challenges in end-to-end testing in a recent survey about the Selenium ecosystem \cite{garcia2020survey}. In this paper, we propose the adoption of fully automated driver management as a solution to mitigate these problems.

\begin{figure}
\centering
\includegraphics[width=0.99\textwidth]{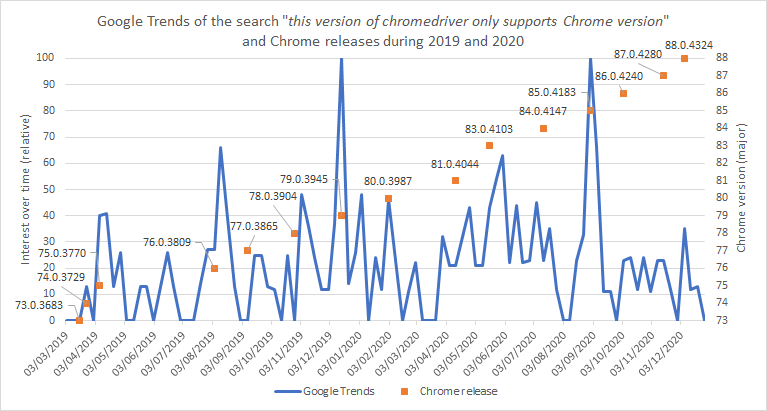}
\caption{Worldwide relative interest of the search term "this version of chromedriver only supports chrome version" in Google Trends together with the release dates of Chrome during 2019 and 2020.}
\label{f3}
\end{figure}

As explained in Section \ref{selweb}, Selenium WebDriver controls web browsers natively thanks to some intermediate binary files called drivers. The management (understood as the process of downloading, configuring, and maintaining) of these drivers is typically carried out manually \cite{garcia2020survey}. This manual process is problematic at several levels. First, it introduces an additional step that impacts the development and maintenance costs of scripts based on Selenium WebDriver. Second, the browser and driver versions are eventually incompatible due to the automated update of modern web browsers (the so-called ``evergreen'' web browsers). In other words, a manually-managed driver Selenium WebDriver test is unreliable (i.e., flaky) due to the unequivocal version mismatch between browsers and drivers in the mid to long term. The users of Chrome and chromedriver experience this problem when a decayed test fails with the error message ``\textit{this version of chromedriver only supports chrome version N}'' (being N the is the minimum version of Chrome supported by a particular version of chromedriver). Figure \ref{f3} shows this fact, since the interest over time of this search term on Google is directly related to Chrome updates incompatible with previous versions of chromedriver \cite{garcia2021automated}.

Overall, driver management is a real issue Selenium WebDriver developers face, and as depicted in Figure \ref{f3}, they actively look for a solution when the incompatibility problem between browser and driver eventually breaks their tests. In a recent survey about driver management in Selenium WebDriver \cite{garcia2021automated}, 76.3\% of respondents declared to adopt WebDriverManager\footnote{\url{https://github.com/bonigarcia/webdrivermanager}} (a Java library for automated driver management) to reduce the development cost of Selenium WebDriver by downloading the required driver automatically. In addition, 69.6\% of participants also declared that WebDriverManager decreases the maintenance costs thanks to the automated update of these drivers. In light of these results, we propose using a fully automated driver management process as a first step in the implementation of the proposed programming model.

Another challenging aspect of end-to-end testing with Selenium reported in the survey published in \cite{garcia2020survey} is the difficulty of developing specific tests, such as load tests. Load testing tools, such as Apache JMeter, have been extensively used to evaluate the performance of web applications. This kind of tool allows sending multiple HTTP requests to a given URL endpoint while measuring the response time \cite{nevedrov2006using}. This approach is not suitable when actual browsers are required to generate the load. Section \ref{cas} provides an example case using an example of these types of applications, i.e., videoconference services based on WebRTC.

The proposed solution to overcome this challenge is the integration of Docker in the Jupiter programming model. Docker is an open-source project that automates the deployment of Linux applications as portable containers \cite{bernstein2014containers}. This technology can support Selenium WebDriver by providing a dockerized infrastructure for different types and versions of web browsers. In addition to load testing, the automated management of web browsers using Docker containers provides other relevant benefits, such as:

\begin{itemize}
  \item Cross-browser testing. Cross-browser compatibility refers to the ability to support web applications on different browsers identically \cite{mesbah2011automated}. Parameterized tests using Selenium WebDriver and Docker containers (i.e., reusing the same test logic with different web browsers) can be used to implement comprehensive cross-browser tests.
  \item Mobile testing. Although Selenium WebDriver is used primarily to verify web applications, it can easily be extended to test mobile applications. To this aim, mobile browsers in Docker containers can be used.
\end{itemize}

The last challenge we aim to solve in this paper is troubleshooting (also known as failure analysis) for Selenium WebDriver. Troubleshooting is the process of gathering and analyzing data to discover the cause of a failure. This process can be complex for Selenium WebDriver tests because the whole system is tested through the User Interface (UI), and the underlying root cause of a failed end-to-end test can be diverse (such as a fault in the client-side, server-side, or the connection with external services, to name a few). To overcome this problem, we propose to bring observability to Selenium WebDriver tests. The term observability comes from the classical control theory by Kalman \cite{kalman1970lectures} and refers to the ability to infer the internal state of a system through the collection and analysis of its external outputs \cite{niedermaier2019observability}. We propose different mechanisms to implement observability in Selenium WebDriver tests. First, we use the Jupiter extension model for monitoring failed tests. Then, we use instrumented Docker containers for gathering data (e.g., session recordings) and configurable browser screenshots when tests fail.

To summarize this section, table \ref{t4} recapitulates the challenges that motivate this work, together with the proposed solution, some implementation details, and the expected benefits.

\begin{table}
\caption{Summary of the end-to-end challenges addressed by Selenium-Jupiter}
\label{t4}
\begin{tabularx}{\textwidth}{L{1.75cm} L{2.2cm} L{3.35cm} L{3.7cm}} \\
\hline\noalign{\smallskip}
\textbf{Challenge} & \textbf{Proposal} & \textbf{Implementation}  & \textbf{Expected benefits} \\
\midrule

Maintainability and flakiness
&
Automated driver management
&
Integration of WebDriverManager into the Jupiter parameter resolution mechanism
&
- Reduce development and maintenance costs \newline
- Reduce flakiness due to incompatible driver \newline
- Reduce test code boilerplate
\\
\midrule

Difficult to develop end-to-end tests
&
Automated infrastructure management
&
Integration of Docker in Jupiter using Java annotations and test templates
&
- Ease cross-browser testing \newline
- Ease mobile testing  \newline
- Ease performance testing
\\
\midrule

Troubleshooting
&
Observability
&
Jupiter extension model for monitoring failed tests plus instrumented Docker containers for data gathering
&
- Ease screenshotting, remote access, and session recording \newline
- Enable configurable browser reporting for failed tests
\\
\midrule

\end{tabularx}
\end{table}

\section{Design}
\label{des}

Selenium-Jupiter is an extension of the Jupiter programming model. Therefore, its design is based on implementing different extension points of the Jupiter extension API (summarized in Table \ref{t1} of Section \ref{back}). Table \ref{t2} shows the extension points used by Selenium-Jupiter. The details of these aspects are explained in the following subsections.

\begin{table}
\caption{Extension points used by Selenium-Jupiter}
\label{t2}
\begin{tabularx}{\textwidth}{L{1.7cm} L{3.8cm} L{5.6cm}} \\
\hline\noalign{\smallskip}
\textbf{Category} & \textbf{Extension point(s)} & \textbf{Main related features} \\
\midrule

Parameter resolution
&
\texttt{ParameterResolver}
&
- \texttt{WebDriver} objects instantiation \newline
- Automated driver management \newline
- Docker containers initialization
\\
\midrule

Test lifecycle
&
\texttt{AfterTestExecution\-Callback}, \texttt{AfterEachCallback}, and \texttt{AfterAllCallback}
&
- Browser disposal \newline
- Stop and remove Docker containers \newline
- Monitoring failed tests \newline
- Gathering browser data (recordings and screenshots)
\\
\midrule

Test templates
&
\texttt{TestTemplateInvoca\-tionContextProvider}
&
- Cross-browser testing

\\
\midrule

\end{tabularx}
\end{table}

\subsection{Parameter resolution}

First, the parameter resolution mechanism provided by Jupiter is used in Selenium-Jupiter. To this aim, Selenium-Jupiter implements the extension point \texttt{ParameterResolver} of the Jupiter extension model. This way, the parameters defined in methods or constructors in test classes are instantiated by Selenium-Jupiter. In particular, Selenium-Jupiter allows injecting instances of the \texttt{WebDriver} class hierarchy to control different types of browsers. Table \ref{t3} shows a summary of the supported types by Selenium-Jupiter, together with the browsers that can be controlled using these types.

When controlling local browsers, Selenium-Jupiter manages the required driver before the instantiation of the declared parameter. Internally, the driver management procedure is done automatically using WebDriverManager. This process follows a driver resolution algorithm made up of several steps. First, the local browser version to be controlled is detected dynamically (i.e., at runtime). Then, this version is used to discover the compatible driver version. Finally, the driver is downloaded from its online repository, is stored locally, and is used to control the browser \cite{garcia2021automated}.

\begin{table}
\caption{Types used for dependency injection in Selenium-Jupiter tests}
\label{t3}
\begin{tabularx}{\textwidth}{L{5.4cm} L{5.6cm}} \\
\hline\noalign{\smallskip}
\textbf{Type} & \textbf{Browser} \\
\midrule

\texttt{ChromeDriver}
&
Google Chrome
\\
\midrule

\texttt{FirefoxDriver}
&
Mozilla Firefox
\\
\midrule

\texttt{EdgeDriver}
&
Microsoft Edge
\\
\midrule

\texttt{OperaDriver}
&
Opera
\\
\midrule

\texttt{SafariDriver}
&
Apple Safari
\\
\midrule

\texttt{ChromiumDriver}
&
Chromium
\\
\midrule

\texttt{InternetExplorerDriver}
&
Microsoft Internet Explorer
\\
\midrule

\texttt{WebDriver} or \texttt{RemoteWebDriver}
&
Remote or dockerized browsers
\\
\midrule

\texttt{List<WebDriver>}
&
List of browsers of a given type 
\\
\midrule

\end{tabularx}
\end{table}

The driver management process is not required when using Docker containers instead of local browsers since the proper driver is already included in the Docker container. In this case, Selenium-Jupiter starts the container images pulled from Docker Hub\footnote{\url{https://hub.docker.com/}}. Selenium-Jupiter uses the following sets of open-source Docker images:

\begin{itemize}
  \item Stable versions of web browsers: Chrome, Firefox, Edge, and Opera. These images are maintained by Aerokube\footnote{\url{https://aerokube.com/images/latest/}}, a company aimed to provide practical solutions for Selenium test infrastructure. These containers are instrumentalized with a Virtual Network Computing (VNC) server that allows monitoring capabilities, such as remote access and session recording.
  \item Beta and development web browsers: Using a fork of the Aerokube images for the beta and development versions of Chrome and Firefox, maintained by the company Twilio\footnote{\url{https://hub.docker.com/r/twilio/selenoid/}}.
\end{itemize}

\subsection{Test lifecycle}

Selenium-Jupiter enhances the test lifecycle by including custom logic after the execution of each test. The extension point \texttt{AfterTestExecution\-Callback} is used for test monitoring. This way, before releasing the resources, Selenium-Jupiter collects session recordings (when using Docker) or makes browser screenshots at the end of tests. These assets can be used for troubleshooting (i.e., failure analysis). Then, the extension point \texttt{AfterEachCallback} is used to release the employed resources. Specifically, local browsers are closed, remote sessions are terminated, and Docker containers are stopped and removed. 

A particular feature of Selenium-Jupiter is called ``single-session''. This feature allows reusing the same browser session by all the tests of a Java class. In this case, the release procedure should be done after executing all tests (and not after each test). For that, Selenium-Jupiter also implements the extension point called \texttt{AfterAllCallback}.

\subsection{Test templates}
\label{43}

The last extension point implemented by Selenium-Jupiter is \texttt{TestTemplate\-InvocationContextProvider}. This extension point allows gathering some context to invoke methods annotated with \texttt{@TestTemplate}. This context is named ``browser scenario'' in the Selenium-Jupiter jargon, and it can be seen as a browser collection to be used with the same test logic. The definition of a browser scenario in Selenium-Jupiter is done in two ways: first, using a custom JSON notation; second, using the Selenium-Jupiter API configuration capabilities. 

\section{Features}
\label{fea}

\begin{figure}
\centering
\includegraphics[width=0.99\textwidth]{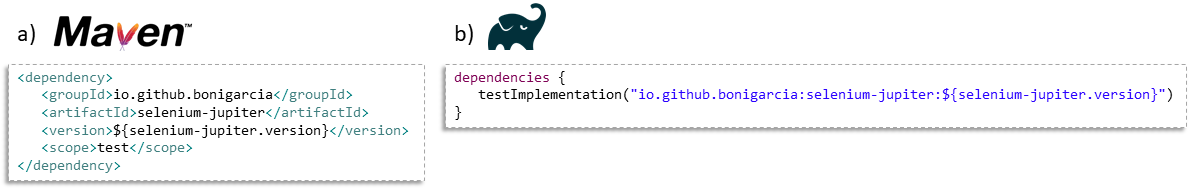}
\caption{Selenium-Jupiter setup in Maven and Gradle}
\label{f4}
\end{figure}

Each release of Selenium-Jupiter is available on the public repository Maven Central\footnote{\url{https://search.maven.org/artifact/io.github.bonigarcia/selenium-jupiter}}. A build tool (such as Maven or Gradle) is typically used to declare and resolve Java dependencies. For that, the project coordinates (\texttt{groupId}, \texttt{artifactId}, and \texttt{version}) are specified. Figure \ref{f4} shows the required configuration to use Selenium-Jupiter by test classes in a Maven and Gradle project. In these snippets, the variable \texttt{selenium-jupiter.version} should be substituted for a given Selenium-Jupiter version (the latest available in Maven central is recommended by default).

The Selenium-Jupiter documentation\footnote{\url{https://bonigarcia.dev/selenium-jupiter/}} provides a complete reference of the Selenium-Jupiter features. The following subsections give a summary of these features.

\subsection{Local browsers}

\begin{figure}
\centering
\includegraphics[width=0.99\textwidth]{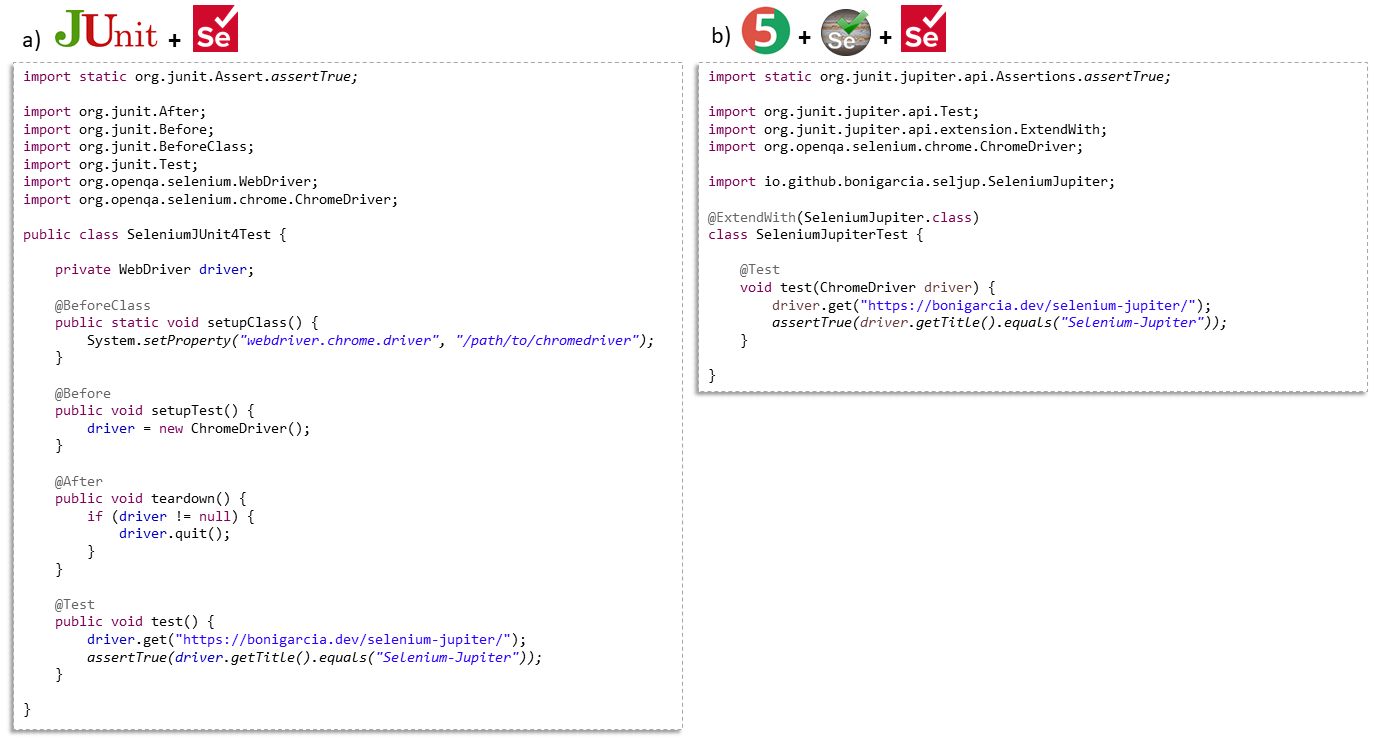}
\caption{Comparison of Selenium WebDriver tests using local web browsers with: a) JUnit 4. b) JUnit 5 and Selenium-Jupiter}
\label{f5}
\end{figure}

Selenium-Jupiter uses an Inversion of Control (IoC) pattern \cite{sobernig2010inversion} for the instantiation of \texttt{WebDriver} objects. In other words, developers using Selenium-Jupiter do not need to create these objects explicitly. Instead, they need to select the WebDriver types injected as test parameters and then use the injected \texttt{WebDriver} instances in the test logic to control web browsers. Moreover, these objects are adequately terminated after each test by Selenium-Jupiter. Figure \ref{f5} shows an example of this feature. On the one hand, Figure \ref{f5}-a uses JUnit 4 and Selenium WebDriver to implement an end-to-end test. As can be seen in this snippet, the driver management and \texttt{WebDriver} instantiation and disposal are done manually in the test setup and teardown. On the other hand, Figure \ref{f5}-b implements the same test logic with Selenium WebDriver, but this time using JUnit 5 and Selenium-Jupiter. As can be seen, tests based on Selenium-Jupiter have a lean implementation compared to a manual approach since the boilerplate to set up and tear down \texttt{WebDriver} instances is not required with Selenium-Jupiter. Boilerplate is the name given to those code sections included repeatedly in different parts of a piece of software with little or no alteration \cite{nam2019marble}.

\subsection{Remote browsers}

Selenium-Jupiter provides a couple of custom annotations to control remote browsers: 

\begin{itemize}
  \item \texttt{@DriverUrl}: URL of the machine hosting the browsers. This URL typically identifies a remote Selenium Hub or a cloud provider endpoint (e.g., Sauce Labs or BrowserStack).
  \item \texttt{@DriverCapabilities}: Key-value pairs that identify the remote browser, such as browser name, browser version, or platform.
\end{itemize}

\begin{figure}
\centering
\includegraphics[width=0.99\textwidth]{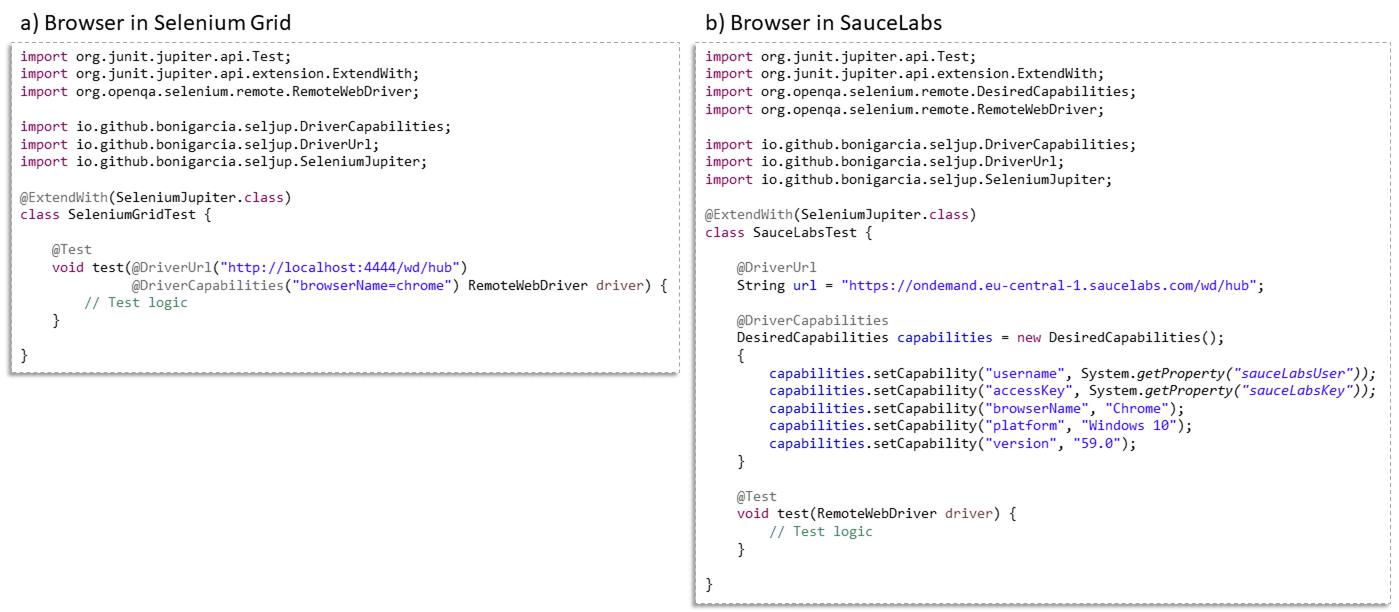}
\caption{Selenium-Jupiter test controlling remote browsers}
\label{f6}
\end{figure}

These annotations are defined at the parameter or field level. Figure \ref{f6} shows an example of this feature. First, Figure \ref{f6}-a shows a test skeleton to control a remote browser using the abovementioned annotations at parameter-level. Then, \ref{f6}-b shows another test skeleton to use a remote web browser on SauceLabs. In this case, the annotations are declared at the field-level.

\subsection{Docker browsers}

Selenium-Jupiter provides the annotation \texttt{@DockerBrowser} to use browsers in Docker containers in Selenium WebDriver tests. This annotation is declared at parameter-level and allows to specify a \texttt{RemoteWebDriver} or \texttt{WebDriver} object used to control web browsers in Docker containers. This feature allows using a wide range of browser types and versions. Furthermore, it enables other advanced testing capabilities, such as session recording, remote access using the VNC protocol, and performance testing (declaring a browser \texttt{List} as test parameter).

\begin{figure}
\centering
\includegraphics[width=0.99\textwidth]{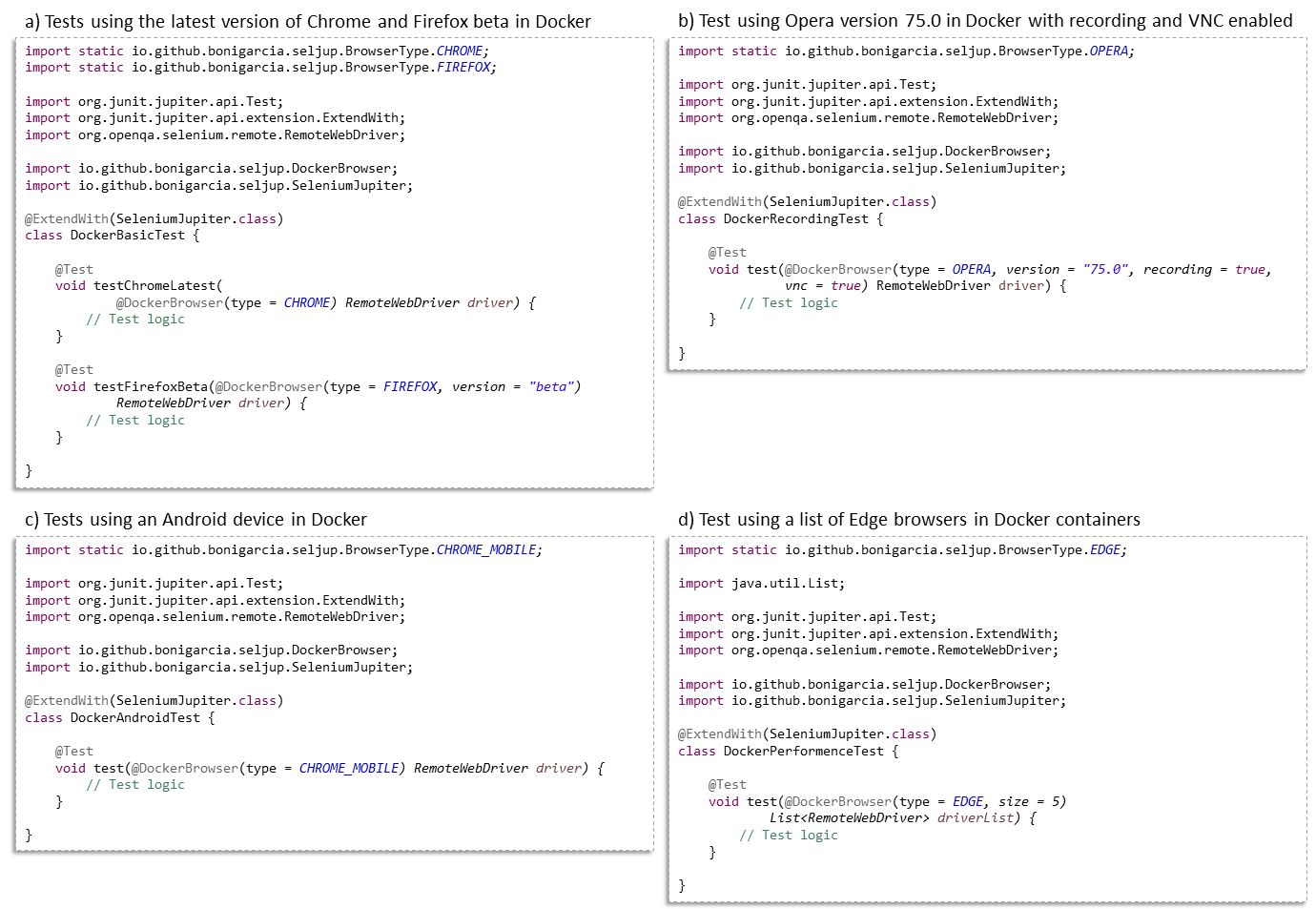}
\caption{Selenium-Jupiter tests using Docker}
\label{f7}
\end{figure}

Figure \ref{f7} illustrates the usage of this feature. The first snippet (Figure \ref{f7}-a) shows a Java class in which two JUnit 5 tests are declared. These tests use the latest version of Chrome and Firefox beta, respectively. Selenium-Jupiter discovers these versions at runtime. To that aim, Selenium-Jupiter requests Docker Hub and selects the proper container image. Figure \ref{f7}-b shows another skeleton example in which a fixed version of Opera is used. Also, Selenium-Jupiter is configured to record and enable remote access to the session with VNC. Next, Figure \ref{f7}-c shows an example in which a Chrome mobile in an  Android device is controlled with Selenium WebDriver. Finally, Figure \ref{f7}-d shows a test example in which a list of Edge browsers is declared.

\subsection{Test templates}

\begin{figure}
\centering
\includegraphics[width=0.99\textwidth]{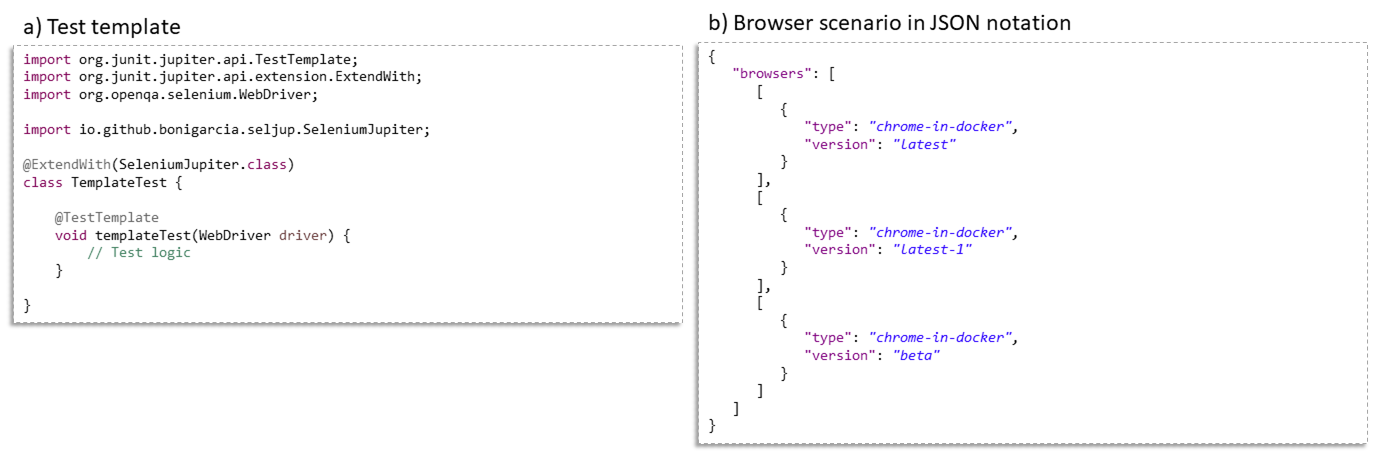}
\caption{Example of test template and browser scenario}
\label{f8}
\end{figure}

Jupiter provides test templates for repeating the same test logic using custom information. As introduced in subsection \ref{43}, this information is given in Selenium-Jupiter using a JSON notation or programmatically. Figure \ref{f8} shows a test template skeleton and browser scenario defined in JSON. In this example, the browser scenario declares three session types, all of them using Chrome in Docker, but with different versions: the stable (\texttt{latest}), the previous version to the stable (\texttt{latest-1}), the beta versions (\texttt{beta}). The logic is repeated three times when this test template is executed, per each browser defined in the scenario.

\subsection{Configuration}

Selenium-Jupiter provides a comprehensive toolbox of configuration parameters aimed to provide a custom setup of the provided features. Each parameter has a single label. For instance, \texttt{sel.jup.recording} enables the recording of Docker sessions, or \texttt{sel.jup.recording.when.failure} allows recording the Docker session only in the case of a failed test. There are three different ways to specify each configuration parameter. The order in which these methods are prioritized is the following:

\begin{enumerate}
  \item Using environment variables. The name for these environmental variables is specified by converting the parameter label (e.g., \texttt{sel.jup.recording}) to uppercase and replacing the dots with underscores (i.e., \texttt{SEL\_JUP\_RECOR\-DING} in the example before).
  \item Using Java properties. In this case, the configuration label is used directly to specify its value, for example, when invoking Maven or Gradle from the command line (e.g., \texttt{mvn test -Dsel.jup.recording=true}).
  \item Using a Java API. Selenium-Jupiter provides a configuration manager that allows setting up all the configuration parameters (Figure \ref{f7}-b shows an example of this configuration type).
\end{enumerate}

\subsection{Other features}

This subsection provides a summary of other features provided by Selenium-Jupiter, namely:

\begin{itemize}
  \item Disabling test. As of version 4, Selenium-Jupiter allows conditionally skipping tests using Java annotations. First, \texttt{@EnabledIfBrowserAvailable} allows disabling a test when a browser (specified as annotation attribute) is not available in the system. Second, \texttt{@EnabledIfDockerAvailable} disables tests when Docker is not installed. Finally, \texttt{@EnabledIfDriverUrl\-Online} allows skipping tests if the Selenium Server remote URL is offline.
  \item Options. Selenium WebDriver options are used to tune different aspects of browsers (e.g., headless and incognito modes, to name a few). Selenium-Jupiter provides a set of Java annotations to configure these options: \texttt{@Options} (used to configure browser-specific options, such as \texttt{ChromeOp\-tions} or \texttt{FirefoxOptions}), \texttt{@Arguments} (to specify browser arguments), \texttt{@Preferences} (to set browser advanced preferences), \texttt{@Binary} (to set the location of the browser binary), and \texttt{@Extensions} (to specify browser extensions).
  \item Single session. Selenium-Jupiter provides a class-level annotation called \texttt{@SingleSession}. It allows reusing the same browser session by all the tests of a given class. In the default behavior,  browsers are initialized before each test and released after each test. \texttt{@SingleSession} changes this behavior: browsers are initialized once at the test suite beginning and disposed at the end of all tests.
  \item Screenshots. Selenium-Jupiter allows making screenshots for browser sessions at the end of each test using configuration capabilities. These screenshots can be encoded as Base64 or stored as PNG images.
  \item Generic driver. Selenium-Jupiter defines a ``generic driver'' as a test parameter of the type \texttt{RemoteWebDriver} or \texttt{WebDriver}. The specific kind of this browser is later established using configuration capabilities.
  \item Custom driver. Selenium-Jupiter allows extending the supported drivers by registering a custom \texttt{WebDriver} object. The type of this custom driver is used later to declare test parameters following the usual programming model.
  \item Integration with Jenkins. The artifacts produced by Selenium-Jupiter (e.g., MP4 recording or PNG screenshots) can be attached to the web interface of the open-source automation server Jenkins\footnote{\url{https://www.jenkins.io/}}. This feature is enabled using configuration capabilities.
\end{itemize}

\section{Example case}
\label{cas}

As of 2020, society has become more and more virtual due to the global pandemic of the COVID-19. Hence, many people have been forced to adapt rapidly to remote work \cite{miller2021your}. WebRTC has become a critical technology in this context since it supports web-based videoconference services using standard technologies. 

Videoconference services based on WebRTC typically use some intermediate infrastructure to support the communication among the participants. SFU (Selective Forwarding Unit) is a commonly adopted architecture for these WebRTC videoconference services.  In SFU, a centralized server (often named ``media server'') receives multiple video and audio streams from the participants, forwarding those streams to the rest of the participants \cite{nevedrov2006using}. This centralized workflow could be a bottleneck if it is not implemented and dimensioned correctly. For this reason, end-to-end performance testing is essential to evaluate the scalability (i.e., the number of participants) and the quality experienced by the final users (QoE).

As introduced in Section \ref{mot}, generating load using HTTP-based tools like Apache JMeter is not suitable for WebRTC since browsers implementing the WebRTC stack are required for recreating the complete user workflow. In contrast, Selenium WebDriver could be an effective tool to carry out end-to-end performance evaluation of WebRTC since it can drive browsers automatically.

This case study provides a practical application of Selenium-Jupiter. The objective is to assess the end-to-end performance of different WebRTC videoconference applications based on SFU. To that aim, Selenium WebDriver and Selenium-Jupiter are used. The research questions (RQs) driving this example case are the following:

\begin{itemize}
  \item RQ1. How can Selenium WebDriver be used to evaluate the end-to-end performance of WebRTC videoconference applications?
  \item RQ2. What are the benefits and limitations of using Selenium-Jupiter to implement these kinds of tests?
\end{itemize}

\subsection{Design}

Nowadays, there are many videoconference systems based on WebRTC available on the Web. For simplicity and replicability, we selected SFU videoconference systems that provide a public online demo. This way, the chosen target applications are the following: Janus\footnote{\url{https://janus.conf.meetecho.com/}}, Jitsi\footnote{\url{https://meet.jit.si/}}, and OpenVidu\footnote{\url{https://openvidu.io/}}.

A test case based on Selenium-Jupiter has been implemented to evaluate each SUT. Each test follows the same structure. First, a local Chrome browser starts (or setup, depending on the SUT) the videoconference. This browser is in charge of compiling the WebRTC statistics (the so-called ``WebRTC stats''\footnote{\url{https://www.w3.org/TR/webrtc-stats/}}). These stats provide a comprehensive collection of performance statistics for each \texttt{RTCPeerConnection} (object of the WebRTC API used to represent the WebRTC connection between two peers) such as bitrate or packets lost, to name a few. The built-in Chrome tool \textit{webrtc-internals} is controlled with Selenium WebDriver to gather this data in the local browser.

Next, the performance feature provided by Selenium-Jupiter is used to generate the rest of the participants. For that, a list of dockerized browsers is used. The list size is a configurable parameter that determines the session load. The Selenium-Jupiter annotation \texttt{@Arguments} is used to specify the following Chrome arguments of each browser (local and dockerized):

\begin{itemize}
  \item \texttt{--use-fake-ui-for-media-stream}: To avoid the need to grant permissions to access the user media.
  \item \texttt{--use-fake-device-for-media-stream}: To feed the user media with a built-in synthetic video provided by Chrome and used for testing.
\end{itemize}

\begin{figure}
\centering
\includegraphics[width=0.8\textwidth]{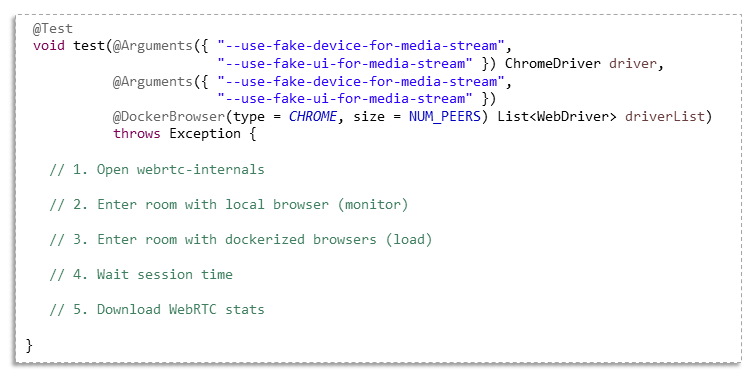}
\caption{Structure of the tests implemented in the example case}
\label{f9}
\end{figure}

Figure \ref{f9} shows the test structure to evaluate each SUT. This figure shows the test method signature in which both the local and dockerized browsers (using the arguments described above) are declared as parameters.  Although this test is specific for WebRTC applications, it illustrates some of the main advantages of using Selenium-Jupiter. First, developers simply declare the type of browser to be controlled with Selenium WebDriver, but they do not deal with the required driver since this is done automatically and transparently. Second, the browser is executed as a Docker container by specifying a Java annotation, hiding the underlying complexity required to use these containers. Finally, Selenium WebDriver tests can use numerous browsers by simply declaring a Java list.

The test logic for the three SUT (i.e., Janus, Jitsi, and  OpenVidu) follows the same procedure, namely:

\begin{enumerate}
  \item Open \textit{webrtc-internals} in a second tab. This way, WebRTC stats are gathered during the session time.
  \item Enter room with local browser. This browser acts as the monitor for the WebRTC session.
  \item Enter room with rest of browsers. These browsers act as the load for the WebRTC session. The selected size for this list is 9, and therefore, the number of browsers used in the tests is 10 (one local plus nine dockerized browsers). This limit is imposed by the free tiers of the evaluated SUTs. Each new browser is connected to the WebRTC session following a configurable time rate. The default value for this rate is 5 seconds, and it allows to simulate users joining a videoconference with a slight time difference.
  \item Wait session time (to simulate conversation with all participants). After checking different periods for the session duration (i.e., 30, 60, and 90 seconds), the selected default value for this session time was 60 seconds since this interval is sufficient to observe the trend of WebRTC indicators.
  \item Download WebRTC stats. The resulting dump file is downloaded using the local browser.
\end{enumerate}

\subsection{Results}

The resulting Selenium-Jupiter tests implemented in this example case have been released in an open-source GitHub repository\footnote{\url{https://github.com/bonigarcia/selenium-jupiter-webrtc}}. A dump file containing the WebRTC stats is obtained after the execution of each test (using Janus, Jitsi, and OpenVidu as SUT). 

Each file dump contains performance stats of the WebRTC peer connection between the local browser (where the data is gathered) and the rest of the browsers (load). The connection with the first dockerized browser allows checking the performance evolution while new peers join the videoconference. For this reason, this connection is analyzed. Among all the performance indicators available in the WebRTC stats, the following parameters of the reception side are chosen:

\begin{itemize}
  \item Bit rate: Total number of bytes received per second.
  \item Jitter delay: Jitter is defined as the latency variation on the packet flow (since some packets might take longer to travel than others). The jitter delay is the sum of time each packet exits in the jitter buffer \cite{garcia2019understanding}.
  \item Freeze count: Total number of video freezes experienced by the receiver. A frame is frozen when the time interval between two consecutively rendered frames equals or exceeds a given threshold. 
  \item Packet loss: Total number of packets lost for a peer connection, calculated as defined in RFC 3550 \cite{schulzrinne2003rfc3550}.
\end{itemize}

\begin{figure}
\centering
\includegraphics[width=0.99\textwidth]{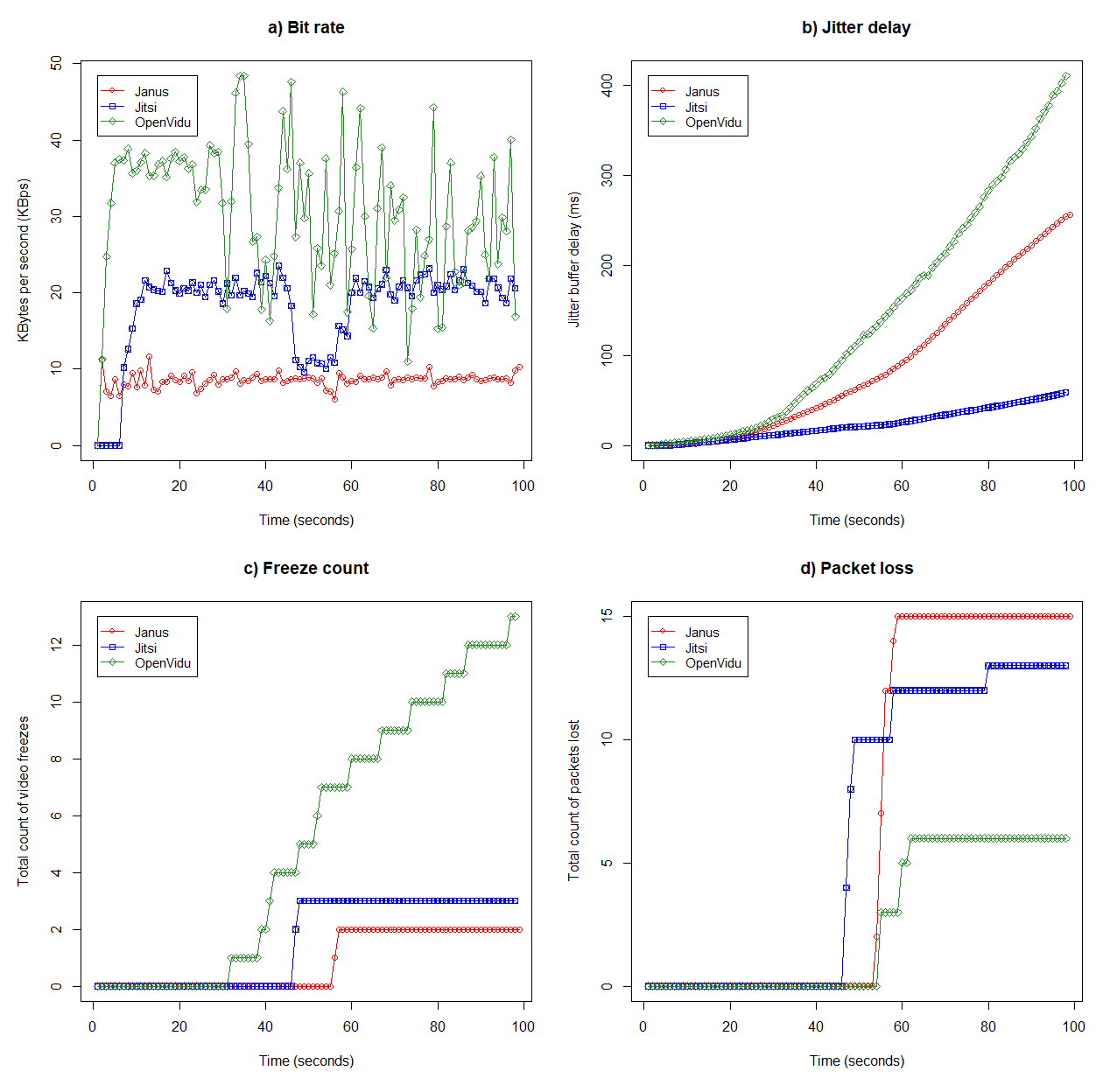}
\caption{Performance results of the connection between the local and the first dockerized browser}
\label{f10}
\end{figure}

We repeated this experiment fives times for each SUT, obtaining equivalent results in each attempt. Figure \ref{f10} shows the results of each parameter during the session time in a single experiment. The participants connect to the WebRTC videoconference during the first part of the session. Since each new participant is connected at a rate of 5 seconds, this part lasts the first 40 seconds of each chart. Then, all the participants are connected to the room for another 60 seconds (i.e., until the second 100 in the charts). 

Figure \ref{f10}-a shows the received bit rate evolution (in KBps). The average is 8.51 KBps in Janus, 18.05 KBps in Jitsi, and 30.52 KBps in OpenVidu. Although the lower average corresponds to Janus, it remains in a more constant range than Jitsi and Janus during experiment time. Figure \ref{f10}-b presents the results regarding jitter delay. This chart shows a monotonically increasing function for all SUTs. As expected, this growth is higher while more participants join the session. At the end of the session, the jitter delay reaches its maximum value: 256.65 ms for Janus, 58.76 ms for Jitsi, and 410.69 ms for OpenVidu. A threshold of 75 ms is recommended to avoid distortions \cite{chong2004comparative}. Jitsi is the only application that does not surpass this threshold in this case study. Figure \ref{f10}-c illustrates the evolution in the freeze count in the video track. On the one hand, both Janus (2 freezes) and Jitsi (3 freezes) experience fewer frozen frames. On the other hand, there are more frozen frames in OpenVidu (a total of 13), and besides, this effect starts earlier (at second 33 in this experiment). Finally, Figure \ref{f10}-d depicts the number of packet losses during the session time. This picture shows that packet loss starts when all participants join the session (i.e., after the second 40). The final count of packets lost is 15 in Janus, 13 in Jitsi, and 6 in OpenVidu.

\subsection{Analysis}

This example case shows a practical application of Selenium-Jupiter with a specific technology: WebRTC. Regarding RQ1, we can conclude that WebRTC stats are a meaningful mechanism to evaluate different performance metrics. These stats can be easily gathered using Selenium WebDriver and the tool \textit{webrtc-internals} available in Chrome.

When coming to RQ2, one of the main benefits of Selenium-Jupiter in the example case is the use of Docker. Thanks to this feature, a developer can choose the type and number of required browsers for end-to-end tests by simply setting selecting an enumerated type and a numeric value. Selenium-Jupiter makes the provisioning of this infrastructure without any need for further setup. The main drawback is related to the scalability of the approach. Since Selenium-Jupiter uses a single Docker engine to handle browsers in containers, it is not feasible to request a large number of containers. In the experiments, using an i7 quad-core laptop with 16 GB of RAM, the maximum number of Chrome browsers in Docker containers is around 100. This figure may be enough for most use cases, although it may be insufficient for more intensive load tests. If more containers are required (e.g., for stress testing), Selenium-Jupiter only escalates vertically, i.e., using more physical resources (RAM and CPU) in a single node.

\section{Discussion}
\label{dis}

This paper contributes to the browser automation space by proposing a programming model on the top of Selenium WebDriver and JUnit 5. This programming model aims to give solutions for several of the relevant challenges in end-to-end testing. 

The presented example case provides hints about the benefits of using Sele\-nium-Jupiter in a particular use case: load testing of WebRTC applications. This example case shows that the integration with Docker can be beneficial for the development of end-to-end tests in different manners. First, it supports cross-browser testing using a rich infrastructure for end-to-end tests, using a wide variety of browsers types and versions in Docker containers. The latest version of the abovementioned container images available in Docker Hub is discovered at runtime. These characteristics ease the creation of advanced test cases since the browser infrastructure is configured and updated automatically. Moreover, this feature can be used to easily implement end-to-end load tests (i.e., when using a large number of browsers).

\subsection{Limitations}

The scope of the experimental case presented in this paper is insufficient to measure the potential advantages of Selenium-Jupiter objectively.

First, the development and maintenance costs promise to be lower using an automated driver management process in Selenium-Jupiter. This fact could be especially relevant in ``evergreen'' web browsers (e.g., Chrome, Firefox, or Edge) since the rapid upgrade rate of these browsers poses a problem for WebDriver tests due to the eventual incompatibility between driver and browser. Nevertheless, the experimental case described in this paper does not prove the development and maintenance costs are lower in Selenium-Jupiter tests. Therefore, this aspect requires further attention in future research.

The second aspect that still needs additional validation is troubleshooting. Selenium-Jupiter allows gathering browser screenshots in case of failure using a rich configuration toolbox (through environmental variables, Java properties, and a Java API). In addition, and thanks to instrumented Docker containers, the browser session can be recorded or accessed remotely. These features are promising, but some experimental research that backs their efficiency is still missing. A possible continuation of this work would include surveying Selenium-Jupiter developers to get a detailed overview of its usage. This survey might help discover the real problems when using Selenium WebDriver and the benefits of using the programming model implemented in Selenium-Jupiter.

\section{Conclusions}
\label{con}

Selenium WebDriver is an open-source library that allows driving web brow\-sers using a language binding. It is used by thousands of companies worldwide to support end-to-end testing. This paper presented Selenium-Jupiter, a JUnit 5 extension for using Selenium WebDriver from Java tests using a comprehensive programming model. The two most relevant innovations shipped with Selenium-Jupiter are automated driver management and seamless integration with Docker. 

The example case presented in this paper illustrates how the Jupiter extension model used in conjunction with instrumented Docker containers facilitates the development of complex test scenarios (e.g., end-to-end load testing for WebRTC applications). Nevertheless, some of the potential benefits of Selenium-Jupiter (such as maintainability, flakiness, and troubleshooting) still require further attention in future research. Finally, scalability is a significant limitation of the presented approach. Selenium-Jupiter uses a single Docker engine to execute containers. For this reason, the number of browsers served as containers is limited to the resources available in the host running the Docker engine. To solve this limitation, we plan to support the use of a Docker cluster (e.g., through Kubernetes) in a future release of Selenium-Jupiter.


\section*{Acknowledgments}

This work has been supported by FEDER/Ministerio de Ciencia, Innovación y Universidades - Agencia Estatal de Investigación through project Smartlet (TIN2017-85179-C3-1-R), and from the eMadrid Network, which is funded by the Madrid Regional Government (Comunidad de Madrid) with grant No. S2018/TCS-4307. This work was also supported in part by the project Massive Geospatial Data Storage and Processing for Intelligent and Sustainable Urban Transportation (MaGIST), funded by the Spanish Agencia Estatal de Investigación (AEI, doi 10.13039/501100011033) under grant PID2019-105221RB-C44.

\bibliography{bgarcia-jss-si-tests-automation}

\end{document}